**Understanding the Instability of the Halide Perovskite CsPbI$_3$ through Temperature-Dependent Structural Analysis**

*Daniel B. Straus\*, Shu Guo, Milinda Abeykoon, and Robert J. Cava\**

Dr. Daniel B. Straus, Dr. Shu Guo, Prof. Robert J. Cava
Department of Chemistry, Princeton University, Princeton, New Jersey 08544, USA
E-mail: dstraus@princeton.edu, rcava@princeton.edu

Dr. Milinda Abeykoon
National Synchrotron Light Source II, Brookhaven National Laboratory, Upton, New York 11973, USA

**Abstract**

Despite the tremendous interest in halide perovskites in solar cells, the reason that the all-inorganic perovskite CsPbI$_3$ is unstable at room temperature remains mysterious. Here single-crystal X-ray diffraction and powder pair distribution function (PDF) measurements characterize bulk perovskite CsPbI$_3$ from 100 to 295 K to elucidate its thermodynamic instability. While Cs occupies a single site from 100 to 150 K, it splits between two sites from 175 to 295 K with the second site having a lower effective coordination number, which along with other structural parameters suggests that Cs rattles in its coordination polyhedron. PDF measurements reveal that on the length scale of the unit cell, the Pb-I octahedra concurrently become greatly distorted, with one of the I-Pb-I angles approaching 82° compared to the ideal 90°. The rattling of Cs, low number of Cs-I contacts, and high degree of octahedral distortion cause the instability of perovskite-phase CsPbI$_3$. These results provide detailed structural information that may help to engineer greater stability of CsPbI$_3$ and other all-inorganic perovskites for use in solar cells.

Halide perovskites exhibit remarkably long carrier lifetimes and diffusion lengths for a highly defective, solution-processable material, allowing for the creation of solar cells with certified efficiencies of 25.2%.[1–5] Unlike in silicon which can be fabricated with defect densities better than one part per million,[6] halide perovskites have a defect-density orders of magnitude greater than silicon and exhibit ion conduction, both of which typically would result in short carrier lifetimes from increased recombination.[7,8] The dominant halide perovskite in solar energy conversion applications is methylammonium lead iodide, an organic-inorganic hybrid material where the A-site cation is the methylammonium ($CH_3NH_3^+$) ion. The structural properties of methylammonium lead iodide have been widely studied.[9,10] However, methylammonium lead iodide and other hybrid perovskites suffer from stability problems in solar cells, which are hypothesized to be related to the volatile organic cation.[11]

To improve the stability of perovskite-based solar cells, one method has been to replace the volatile organic cation with cesium.[12] While cesium and cesium halides are significantly less volatile than their organic analogues, black, perovskite-phase cesium lead iodide is not stable at room temperature and rapidly converts to a non-perovskite, wider band gap yellow phase that is not useful for solar energy conversion.[13–17] While perovskite-type $CsPbI_3$ has been stabilized using small-grain sizes and nanocrystals,[18,19] its inherent thermal instability has made detailed structural studies of the bulk material difficult to conduct. Furthermore, many tolerance factor models predict that perovskite-phase $CsPbI_3$ is stable, adding to the mystery and suggesting that the relatively small size of the $Cs^+$ cation is not the only factor that causes the instability of the perovskite phase.[20–22] We recently synthesized kinetically stable single crystals of perovskite-phase $CsPbI_3$,[15] allowing us to characterize its bulk properties.

Here, we use temperature-dependent single-crystal X-ray diffraction (SCXRD) measurements as well as powder X-ray diffraction pair distribution function (PDF) measurements[23] to structurally characterize the long-range and local structures of bulk perovskite-type $CsPbI_3$. Between 100 and 295 K, the $CsPbI_3$ perovskite remains orthorhombic in the *Pnma* (#62) space group (γ-CsPbI$_3$) with the GdFeO$_3$ structure type. Notably, from 175-295 K the Cs cation is disordered and occupies two sites in a thermally-activated process, with the occupancy of the second site reaching 18(2)% at 295 K. This behavior suggests the Cs atom rattles within the structure.[24–27] The second Cs exhibits lower coordination to the surrounding I anions, which in accordance with Goldschmidt's principles[20,21] contributes to the instability of CsPbI$_3$ at room temperature.

Like the long-range structure, the local structure on the length scale of the unit cell remains orthorhombic in the *Pnma* space group between 100 and 295 K. However, it exhibits additional structural distortion, which is most clearly seen through deformed Pb-I octahedra. The local distortion is similar to what has been found in organic-inorganic hybrid halide perovskites,[9,28] indicating that local distortions are a common feature of halide perovskites and are not solely caused by the presence of the anisotropic organic cation.

Our study indicates that the rattling of the Cs cation, lower coordination of the second Cs site, and local octahedral distortion contribute to the thermodynamic instability of γ-CsPbI$_3$. These results explain the metastability of perovskite-type γ-CsPbI$_3$ and provide detailed structural information that may help to engineer greater stability of CsPbI$_3$ and other all-inorganic perovskites.

As previously reported,[14,15,17] at 295 K perovskite-type γ-CsPbI$_3$ displays the orthorhombic *Pnma* (#62) space group. Here, we observe no change in symmetry between 100 K and 295 K (Figure S1 and Tables S1-2). **Figure 1**A shows the pseudocubic cell constants as a function of temperature with the unit cell volume in Figure S2. While the unit cell volume, *b*, and *c* increase with temperature, *a* decreases to eventually become metrically equal to *c* upon the phase transition to tetragonal symmetry around 457 K.[17] Common to the perovskite structure type,[29] γ-CsPbI$_3$ becomes less distorted as the temperature increases, which can be seen through the reduction of the Glazer tilt angles a$^+$ and b$^-$ as well as the angles θ, φ, and Φ defined by Zhao et al. (Figure 1B);[30,31] θ and a$^+$ are equal. While the average thermal parameter $U_{eq}$ of Pb varies linearly with temperature, $U_{eq}$ for Cs and I are superlinear above 200 K (Figure 1C). The thermal parameters of iodine are much more oblate than those of Pb (Figures 1D and S3), indicating that the Pb-I bond is rigid and the I atoms predominantly move orthogonally to the Pb-I bond than along it at all temperatures from 100 K to 295 K.[32] Interestingly, the Pb-I bond lengths are invariant with temperature (Figure 1E). This indicates that the change in lattice parameters must come from tilting between adjacent octahedra and/or from angular changes within the octahedra themselves (Figures 1B, 1F).

The most noteworthy change is that as the temperature increases, Cs becomes disordered between two sites. Figure 2A shows the partial Cs density at 100 K, indicating that the Cs atom is well-behaved and a single site is occupied. As the temperature increases above 150 K, however, the Cs scattering density becomes irregular. The partial Cs density at 295 K is shown in Figure 2B and exhibits significant elongation, indicating that Cs is not well-behaved. We find the residual density is minimized by splitting the Cs occupancy between two sites that are within about an angstrom of each other. We call the dominant site $Cs_A$ and the other $Cs_B$. In the structure refinement, the atomic displacement parameters of $Cs_A$ and $Cs_B$ are constrained to be identical and the sum of the occupancies is fixed at 1. If the Cs atom is modeled without disorder, $U_{eq}$ for Cs is significantly more superlinear at temperatures ≥ 200 K (Figure S4) and there is an unacceptable amount of residual scattering density near Cs. We fit the occupancy as a function of temperature to a thermally activated (Boltzmann) model where $\text{Occ}(Cs_B)/\text{Occ}(Cs_A) = \exp(-\Delta E/kT)$ and find that $\Delta E = 46.8(2)$ meV. This model assumes nothing changes except the temperature. However, lattice expansion is also expected to contribute to the split position because the Cs-I coordination polyhedron expands in volume from 198.3 Å$^3$ at 100 K to 202.9 Å$^3$ at 295 K (Figure S5). The elongated Cs scattering density may also be consistent with the existence of anharmonic vibrations, which have been reported in all-inorganic and hybrid lead halide perovskites including CsPbI$_3$.[17,32–34]

The Cs-I coordination polyhedra at 295 K for both $Cs_A$ and $Cs_B$ are shown in **Figure 3**A-B with a histogram of Cs-I distances in Figure 3E. We find that $Cs_B$ is more centrally located in the polyhedron than $Cs_A$, a factor that may influence its stability there. For a perovskite structure to form, the cations must not rattle in the structure;[21] Goldschmidt postulated that cations must contact as many anions as possible for a structure to be stable.[20] The observed distribution of Cs over two sites in the perovskite cavity can be straightforwardly interpreted as due to Cs rattling within its cage of I atoms at temperatures ≥175 K,[24–27] contributing to the thermodynamic instability of γ-$CsPbI_3$.

We hypothesize that in accordance with Goldschmidt's principle of maximum anion contact,[20] the number of good Cs-I contacts determines the stability of the perovskite phase of $CsPbI_3$. A histogram of Cs-I distances is shown in Figure 3E, illustrating the large variance in Cs-I distances for $Cs_A$ and $Cs_B$. We use the Effective Coordination Number (ECoN) formalism[35–37], where $ECoN = \sum_i w_i$ where $w_i = exp\left[1 - \left(\frac{l_i}{l_{av}}\right)^6\right]$ is the $i$th bond's weight, and $l_{av}$ is the weighted average bond length within the coordination polyhedron as defined in ref. [37]. We also compute the distortion index[37,38] $D = \frac{1}{n}\sum_{i=1}^{n}\frac{|l_i - l_{mean}|}{l_{mean}}$ which quantifies the normalized deviation from the average bond length within the polyhedron, where $l_{mean}$ is the unweighted average bond length within the coordination polyhedron. We find the ECoN of $Cs_A$ to be 8.1 and of $Cs_B$ to be 5.9, with $D$ = 0.12 and 0.099 respectively. The standard deviation σ of Cs-I bond lengths for $Cs_A$ is 0.61 Å and for $Cs_B$ is 0.51 Å. In comparison, the ECoN of the Cs-I coordination polyhedron of non-perovskite δ-$CsPbI_3$ (Figure 3C) is larger with a value of 8.9. The polyhedron is also significantly less distorted with $D$ = 0.016 and σ = 0.08 Å. The additional regularity of the Cs coordination environment in δ-$CsPbI_3$ can also be seen in Figure 3E, where the Cs-I contacts are all within 0.25 Å. At room temperature, perovskite γ-$CsPbI_3$ is only metastable and δ-$CsPbI_3$ is the thermodynamically stable phase,[14,15] consistent with non-perovskite δ-$CsPbI_3$ having a significantly less distorted coordination polyhedron with an ECoN 0.8 larger than $Cs_A$ and 3.0 larger than $Cs_B$. The order of magnitude higher $D$ and $\sigma$ values for both $Cs_A$ and $Cs_B$ in perovskite γ-$CsPbI_3$ compared to stable yellow δ-$CsPbI_3$ further support the hypothesis that the $Cs^+$ ion rattles in its coordination polyhedron.

At temperatures >600 K, δ-$CsPbI_3$ spontaneously converts to the cubic perovskite α-$CsPbI_3$, where Cs is in the middle of the iodine cavity.[17,39] Its Cs-I coordination polyhedron is shown in Figure 3D, and there are 12 identical Cs-I contacts resulting in an ECoN of 12 and $D$ and $\sigma$ of 0 (Figure 3E). We believe α-$CsPbI_3$ becomes the thermodynamically stable phase at high temperature because increased thermal motion of the $Cs^+$ and $I^-$ ions allows the longer 4.45 Å Cs-I contacts to be stable while also maximizing the ECoN. Changes in the Pb-I bonds likely do not contribute to the stability of α-$CsPbI_3$ at high temperatures because at 634 K, the Pb-I bond length is 3.14 Å, which only differs by 0.03-0.06 Å from the Pb-I bond lengths we measure in γ-$CsPbI_3$ from 100 K to 295 K.[39]

Our structural study also suggests that perovskite γ-$CsPbI_3$ exhibits increased stability at temperatures below 175 K, where the Cs ion only occupies the $Cs_A$ site, which has a much larger ECoN than $Cs_B$. This hypothesis is supported by the observation that the Cs disorder is a thermally activated process. Using density-functional theory calculations, Marronnier et al. found that γ-$CsPbI_3$ has a lower total energy than non-perovskite δ-$CsPbI_3$, contrary to the observation that δ-$CsPbI_3$ is what forms when $CsPbI_3$ is synthesized at room temperature.[17] It may be possible that γ-$CsPbI_3$ is the thermodynamic product at temperatures ≤175 K, where the Cs atom is well-behaved and occupies only a single site, although the ECoN at 100 K (8.0) does not significantly change from 295 K and is still smaller than the ECoN for yellow δ-$CsPbI_3$ (8.9).

$CsPbI_3$ reveals the limitations of tolerance factor approaches to predicting perovskite stability, especially in systems like the current one. Using the ionic radii from Shannon's table,[40] the Goldschmidt tolerance factor[20,21] $t = \frac{r_A + r_X}{\sqrt{2}(r_B + r_X)}$ for CsPbI$_3$ is 0.851, similar to that of GdFeO$_3$ (0.848),[22] yet the thermodynamically-stable phase of CsPbI$_3$ at room temperature is the non-perovskite δ-CsPbI$_3$. Similarly, based on the observed tolerance factor defined by Sasaki et al. $\tau = \langle A - X \rangle / (\sqrt{2} \langle B - X \rangle)$,[22] where A—X and B—X refer to average bond distances within each coordination polyhedron, the value for γ-CsPbI$_3$ (0.987) is in the range of many stable oxide perovskites, such as CaTiO$_3$ (0.996) and GdFeO$_3$ (0.977). Filip and Giustino incorporate the Goldschmidt tolerance factor and the octahedral factor ($\mu = r_B/r_X$) to find a stable region for perovskites,[21] yet their model predicts CsPbI$_3$ to be stable as a perovskite. Bartel et al. propose a tolerance factor $\tau = \frac{r_X}{r_B} - n_A \left( n_A - \frac{r_A/r_B}{\ln(r_A/r_B)} \right)$ where n$_A$ is the oxidation state of the A-site cation (Cs in CsPbI$_3$), and under their model, a perovskite structure is stable for $\tau < 4.18$.[41] Their model then correctly predicts that CsPbI$_3$ will not be a perovskite at room temperature because $\tau = 4.30$.

In contrast to SCXRD and conventional powder diffraction measurements, PDF measurements take the diffuse scattering into account in addition to the Bragg reflections to generate a pair distribution function $G(r)$, which quantifies the probability of finding two atoms separated by a distance *r*. A major advantage to PDF measurements is that they allow the local structure to be analyzed separately from the long-range average structure.[9,23,28] We fit G(r) with a single Cs site at all temperatures because when attempting to fit G(r) with disorder in the Cs positions, the resulting structure is not physically reasonable. The inability to resolve the long-range disorder we observe in SCXRD measurements is likely due to the relative insensitivity of the PDF method to the kind of disorder we observe in the long-range structure;[42,43] peaks corresponding to the two Cs sites will overlap significantly in the one-dimensional PDF dataset while the three-dimensional dataset provided by SCXRD measurements allows the two sites to be individually resolved.

Like the SCXRD data, we fit $G(r)$ using a structural model with orthorhombic symmetry in the *Pnma* space group. **Figure 4**A-B shows good agreement between the long-range data (r = 20 − 50 Å) and the fit at both 100 K and 295 K; data and fits at intermediate temperatures are shown in Figure S6. However, the structure found by fitting the long-range data does not adequately fit the local region from r = 2.5 − 8 Å, which is on the length scale of a single unit cell unit. Most of the poor correspondence between the long-range fit and the local $G(r)$ is for distances corresponding to Cs-I and I-I bonds.

Fitting the short-range local crystal structure separately, we find much better agreement between $G(r)$ and the fit (Figure 4C-D). At temperatures >175 K, the Cs-I bonds begin to elongate (Figure 4E) and the Pb-I octahedra become highly distorted on the length scale of the unit cell with one of the I-Pb-I angles approaching 82° compared to the ideal 90° (Figure 4F), contributing to the instability of perovskite-type γ-CsPbI$_3$. Because these distortions greatly increase at the same temperatures that the Cs scattering density splits in SCXRD measurements, the increased local octahedral distortion found from PDF measurements and split Cs density observed in the SCXRD data are likely manifestations of the same phenomenon. While an I-Pb-I angle near 82° is drastically different than what is observed in the SCXRD data, we note that at 295 K the equivalent isotropic thermal parameters $U_{eq}$ for I are 2-4x larger in the long-range SCXRD structures than in the local PDF structures while $U_{eq}$ for Pb only differs by ~20% (Figure S7). This is consistent with the larger thermal parameters of I in the long-range structure reflecting the average of all orientations of the distorted local structure, while Pb remains fixed at the center of the octahedra in both the local and long-range structures so local distortions do not significantly affect its behavior.

Like what we find here for perovskite-type γ-CsPbI$_3$, PDF measurements on organic-organic hybrid perovskites such as methylammonium lead bromide and methylammonium lead iodide exhibit different local and long-range structures,[9,28] which has been attributed predominantly to the behavior of the organic cation.[28] Here, the existence of differences between the local and long-range structures in the all-inorganic perovskite γ-CsPbI$_3$ suggests that this behavior does not depend exclusively on the organic cation and is a common feature of halide perovskites. It has also been hypothesized that organic-inorganic hybrid perovskites may exhibit ferroelectricity or the Rashba effect, both of which require non-centrosymmetric structures.[44–46] Here, we find that both the local and long-range structures are centrosymmetric for perovskite-type γ-CsPbI$_3$. This observation rules out the existence of the static Rashba effect in γ-CsPbI$_3$. In contrast, even though the long-range structure in the hybrid perovskite methylammonium lead iodide at 350 K is cubic in the centrosymmetric *Pm-3m* space group, the local structure is reported as being non-centrosymmetric in the tetragonal *I4cm* space group, allowing a conventional Rashba effect to exist.[9] It may still be possible that vibrations in γ-CsPbI$_3$ can result in subtle transient local symmetry breaking, which may allow for a dynamic Rashba effect.[47]

We use SCXRD and PDF measurements to analyze the temperature-dependent structural behavior of orthorhombic perovskite γ-CsPbI$_3$. We find that γ-CsPbI$_3$ at ambient temperature has fewer good Cs-I contacts than non-perovskite δ-CsPbI$_3$, which in accordance with Goldschmidt's principle of maximum anion contact[20] we hypothesize is one reason that yellow δ-CsPbI$_3$ is the thermodynamically stable product at room temperature. Furthermore, between 175 and 295 K we find that the Cs atom is split between two sites in a thermally activated process, which suggests the Cs atom rattles in its cage of I atoms and may contribute to the thermodynamic instability of perovskite-type γ-CsPbI$_3$. The thermally activated Cs site has fewer anion contacts than the dominant site, which may further decrease the stability of γ-CsPbI$_3$ at these temperatures. Between 100 and 150 K, the Cs atom occupies a single site with well-behaved electron density, indicating that γ-CsPbI$_3$ is more stable at these temperatures. PDF measurements show that the local structure of γ-CsPbI$_3$ is more distorted than the global structure between 100 and 295 K, which suggests that the existence of differing local and long-range structures is common to halide perovskites and is not caused by the presence of an anisotropic organic cation.[9,28] The tremendous distortion of the Pb-I octahedra at short range likely also contributes to the instability the γ-CsPbI$_3$ perovskite. Our detailed characterization of the local and long-range structures provides vital structural information that will guide theoretical and experimental studies that aim to stabilize bulk forms of perovskite-phase γ-CsPbI$_3$, allowing them to be used in stable solar cells.

**Experimental**

γ-CsPbI$_3$ is crystallized from the melt using a previously published procedure.[15] A stoichiometric quantity of CsI (Sigma-Aldrich, anhydrous, 99.999%) and PbI$_2$ (Alfa-Aesar, ultra-dry, 99.999%) are sealed in an evacuated quartz ampoule. The ampoule is heated to 550 °C and slowly cooled to 370-400 °C at which point it is quenched in an ice-water bath.

Single-crystal studies are performed on a Bruker Kappa Apex2 CCD Diffractometer using graphite-monochromated Mo-Kα radiation (λ = 0.71073 Å). An Oxford Cryostream 700 cryocooler flows dry nitrogen over the crystal. Background, polarization, Lorentz-factor, and multi-scan absorption corrections are used when processing the data. The initial structure solution is found using the intrinsic phasing method of the ShelXT program[48] and refined using the least-squares algorithm in the ShelXL program[49] using the Olex2 GUI.[50] The 295 K structure presented here is a re-refinement of data from Ref. [15]. Structures are visualized using VESTA.[37] The partial Cs density (Figure 2A-B) is obtained by starting with the refined SCXRD structure, removing Cs, and running a ShelXL least squares refinement cycle with the number of least squares cycles set to 0 (L.S. 0 instruction), which computes structure factors but does not change the atomic positions or displacement parameters.[49] The partial Cs density is then found by subtracting F(calc) from F(obs). We note that the partial Cs density is unchanged whether the Cs atom is modeled with or without disorder in the starting structure.

To perform X-ray PDF measurements, γ-CsPbI$_3$ is ground using a mortar and pestle in an argon-filled glove box and sifted using a 200-mesh sieve. The powder is sealed in evacuated quartz capillaries under high vacuum (~$10^{-4}$ torr). Data are collected at beamline 28-ID-1 at NSLS-II at Brookhaven National Laboratory using 75 keV radiation and are analyzed using the Diffpy-CMI software suite.[51] The PDF Gaussian envelope damping parameter Qdamp and the PDF peak broadening parameter Qbroad are determined from measurements of a Ni standard and are fixed at these values for all analysis of $\gamma$-$CsPbI_3$. At each temperature, the long-range structure ($r = 20 - 50$ Å) from the PDF is fit using the SCXRD structure as the initial guess. All symmetry-allowed positions and thermal parameters as well as the dataset scale factor are varied in this fit. After the long-range structure is determined, the peak sharpening parameter $\delta_1$ used in the short-range fits is determined by fixing all parameters, extending the fit-range to $r = 2.5 - 50$ Å, and fitting $\delta_1$, which is not varied in the short-range fits. The short-range structure is fit from $r = 2.5 - 8$ Å. In the short-range fits, the dataset scale factor is fixed to the value determined from the long-range fit, and isotropic thermal parameters are used.

Further details of the crystal structure investigation(s) may be obtained from the Fachinformationszentrum Karlsruhe, D-76344 Eggenstein-Leopoldshafen (Germany), on quoting the depository number CSD 1984123-1984140.

**Acknowledgements**
The synthesis of the compound and analysis of the diffraction data was supported by the Gordon and Betty Moore Foundation, grant GBMF-4412. The single crystal X-ray diffraction data collection was supported by the U.S. Department of Energy, Division of Basic Energy Sciences, grant DE-SC0019331. This research used resources of the National Synchrotron Light Source II, a U.S. Department of Energy (DOE) Office of Science User Facility operated for the DOE Office of Science by Brookhaven National Laboratory under Contract No. DE-SC0012704.

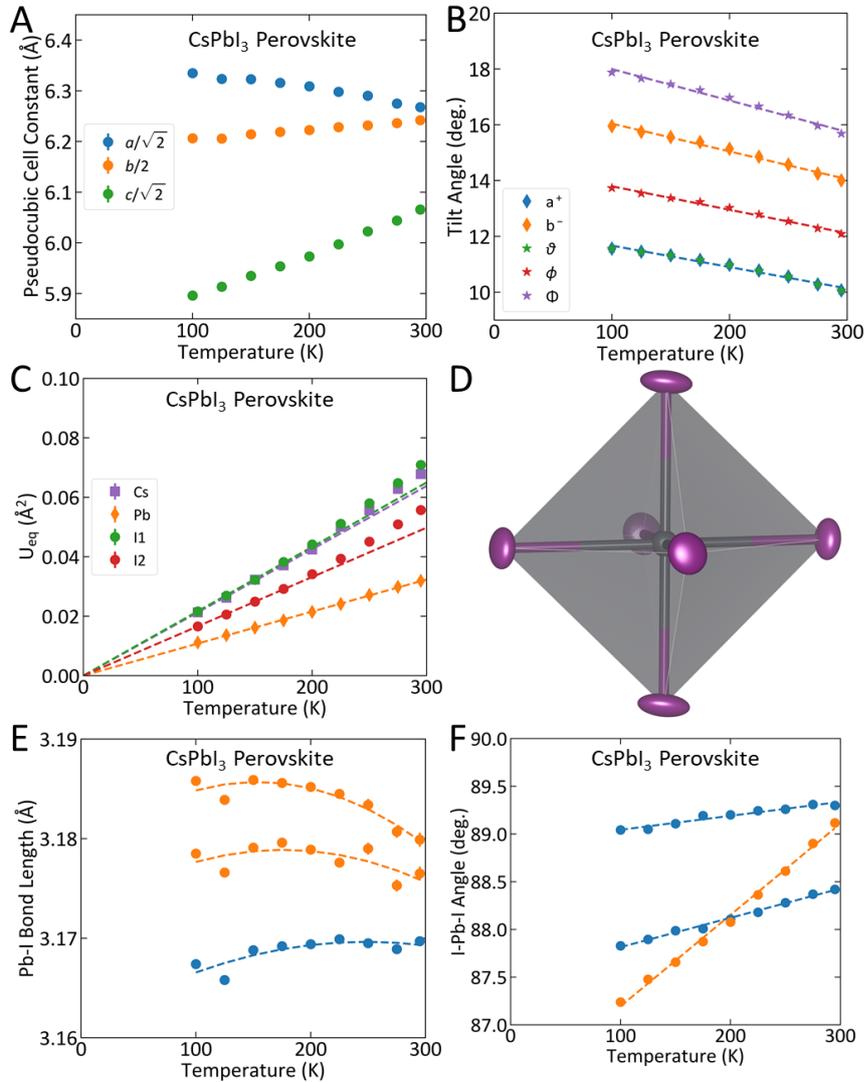

**Figure 1: General characterization of the long-range crystal structure of perovskite-type γ-CsPbI₃.** (A) Normalized cell constants and (B) tilt angles from the SCXRD structures with (dashed lines) linear fits to guide the eye. (C) $U_{eq}$ for Cs (purple squares), Pb (orange diamonds), I1 (green circles), and I2 (red circles), with (dashed lines) linear fits through the first four points. (D) Pb-I octahedron from the 295 K crystal structure. (E) Pb-I bond lengths, with (blue) Pb-I1 and (orange) Pb-I2. (F) I-Pb-I bond angles, with (blue) I1-Pb-I2 and (orange) I2-Pb-I2.

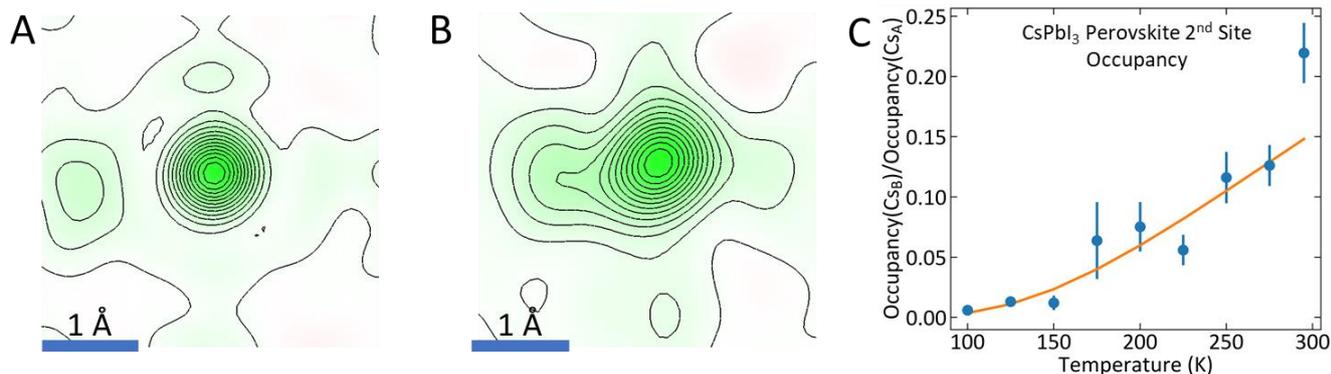

**Figure 2: Structural Disorder at the Cs position in the perovskite phase.** Partial Cs scattering density at (A) 100 K and (B) 295 K. Scale bars are 1 Å. (C) Ratio of the site occupancies of $Cs_B$ to $Cs_A$ as a function of temperature with (orange) Boltzmann fit.

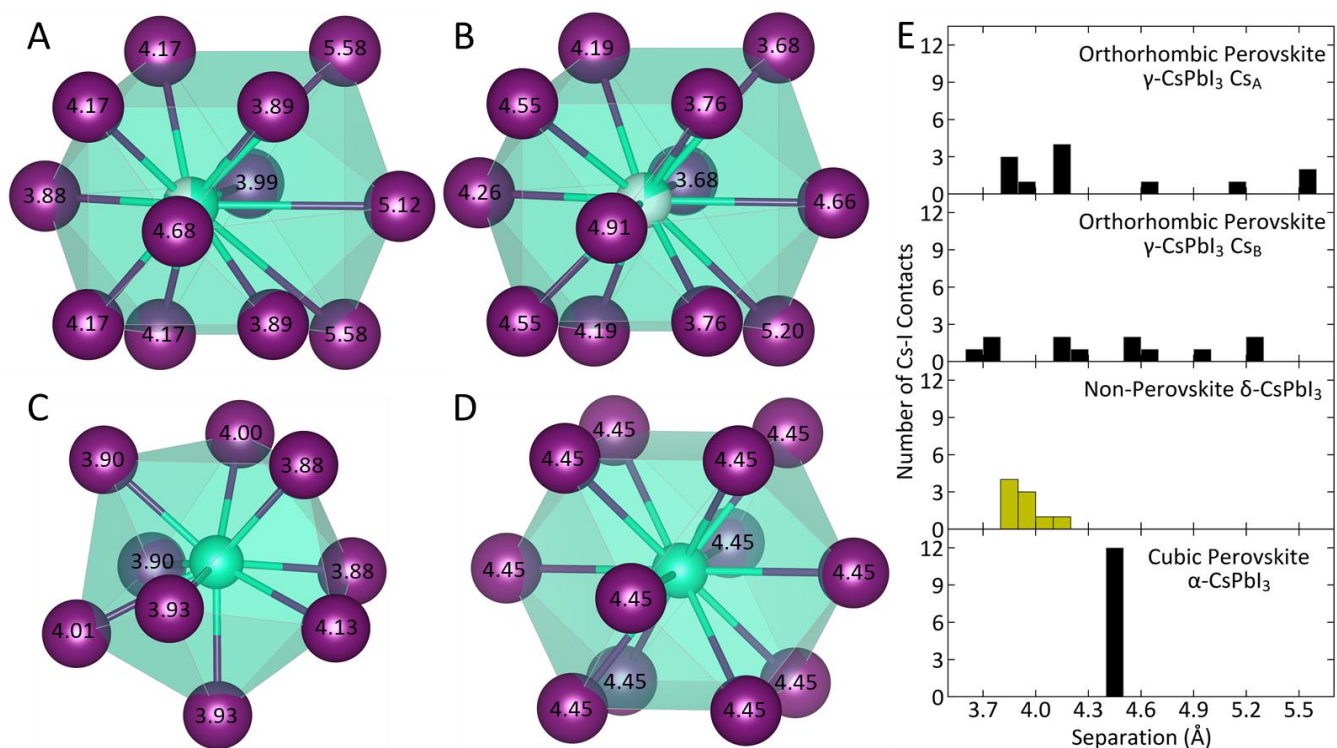

**Figure 3: Comparison of the Cs-I coordination polyhedra for different forms of $CsPbI_3$.** Cs-I coordination polyhedral for (A) $Cs_A$ and (B) $Cs_B$ in the orthorhombic perovskite γ-$CsPbI_3$, (C) for non-perovskite δ-$CsPbI_3$, and (D) for the cubic perovskite α-$CsPbI_3$. (E) Histogram of Cs-I contacts. Structure in (C) from Ref. [15]. Structure in (D) from Ref. [39].

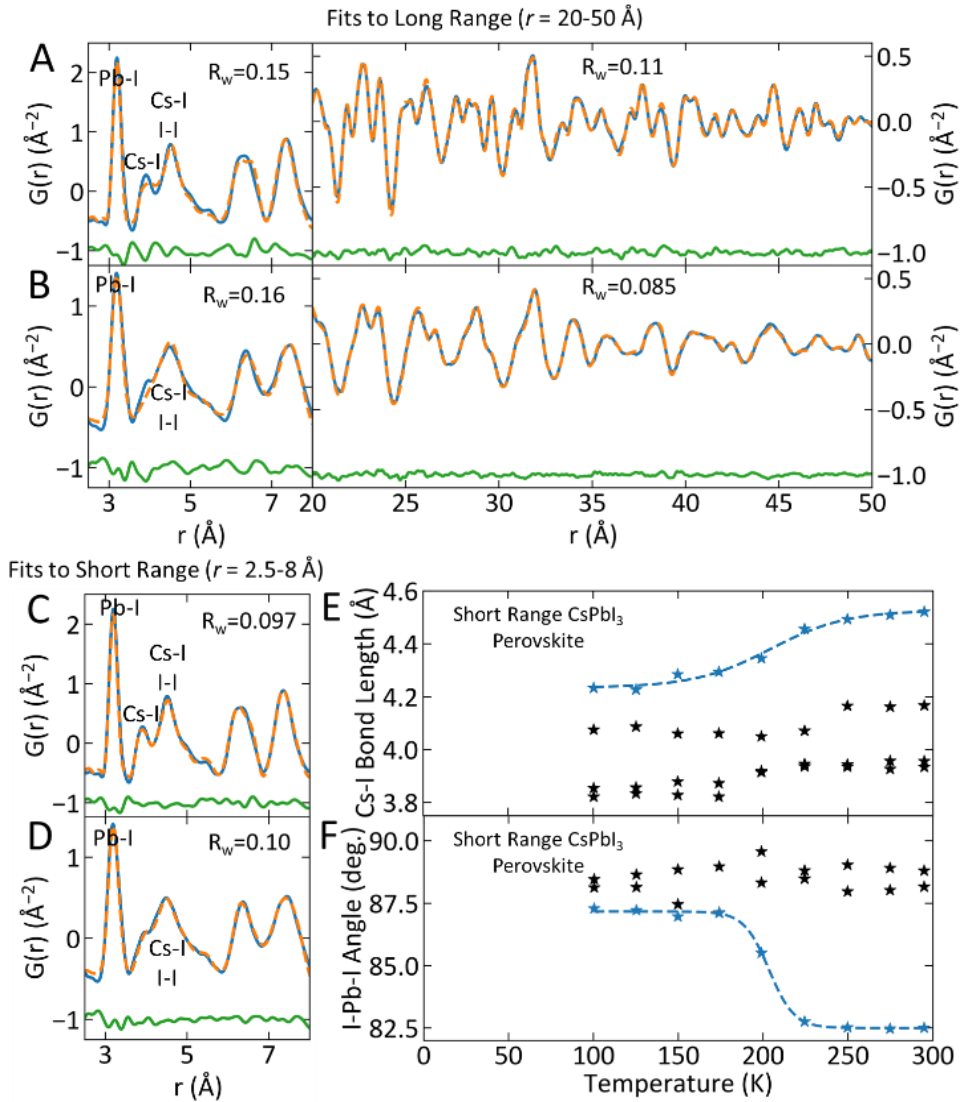

**Figure 4: The local crystal structure of perovskite-type γ-CsPbI₃.** (Blue) G(r) at (A) 100 K and (B) 295 K, with (orange) fit to long-range ($r = 20 - 50$ Å) data and (green) residual. (Blue) G(r) at (C) 100 K and (D) 295 K, with (orange) fit to short-range ($r = 2.5 - 8$ Å) data and (green) residual. (E) Cs-I bond lengths and (F) I-Pb-I angles at short range from PDF fits. Dashed lines are to guide the eye.

# Supporting Information

**Understanding the Instability of the Halide Perovskite CsPbI3 through Temperature-Dependent Structural Analysis**

*Daniel B. Straus\*, Shu Guo, Milinda Abeykoon, and Robert J. Cava\**

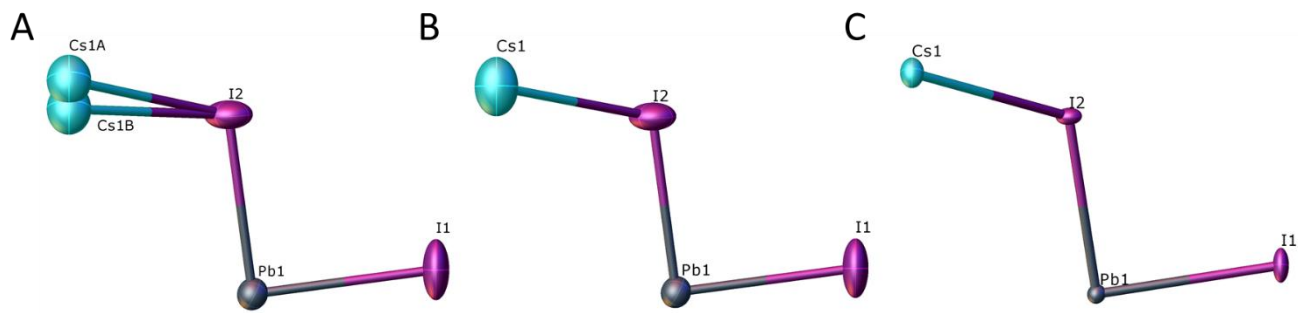

Figure S1: Asymmetric unit of perovskite-type γ-CsPbI$_3$ with 50% probability thermal ellipsoids at (A) 295 K with disordered Cs, (B) 295 K without disordered Cs, (C) 100 K without disordered Cs.

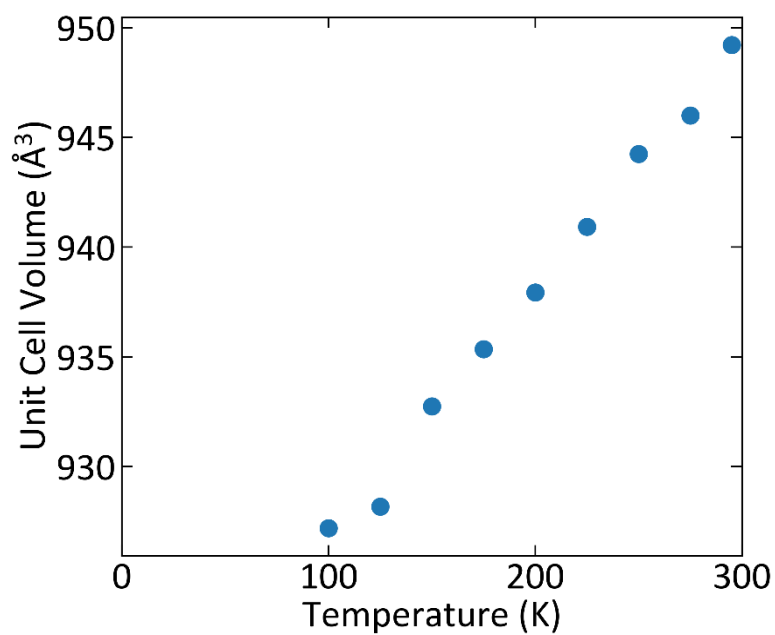

Figure S2: Unit cell volume of SCXRD structures.

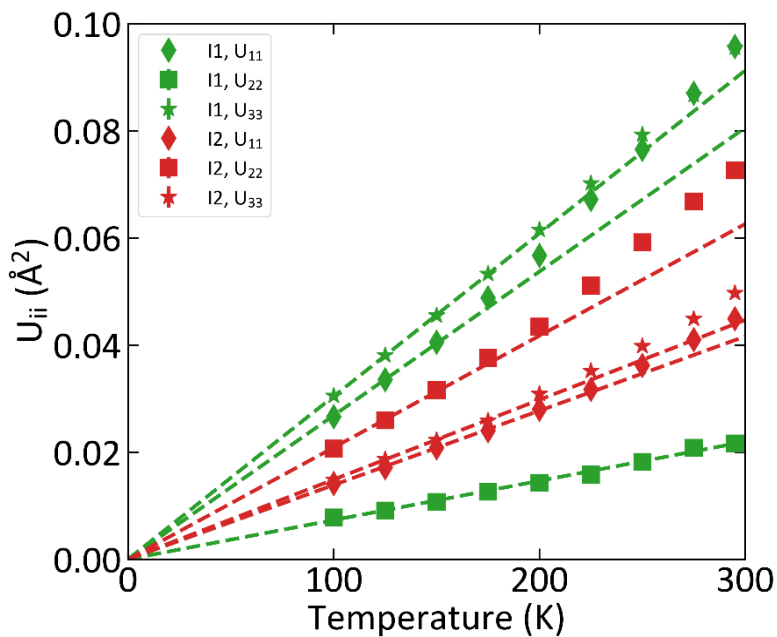

Figure S3: Principal ellipsoids of I1 and I2 from SCXRD structures. Dashed lines are linear fits through the first three points. Within error, the parameters extrapolate to 0 at 0 K, indicating that static displacements of the atoms from their determined positions do not have a significant influence on the thermal parameters.

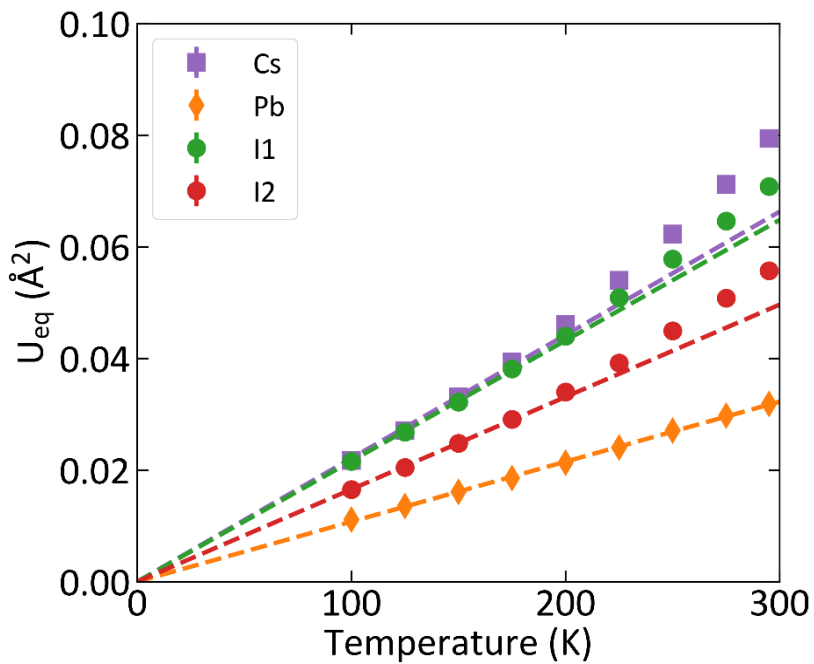

Figure S4: Equivalent isotropic thermal parameters ($U_{eq}$) from SCXRD structures refined without Cs disorder.

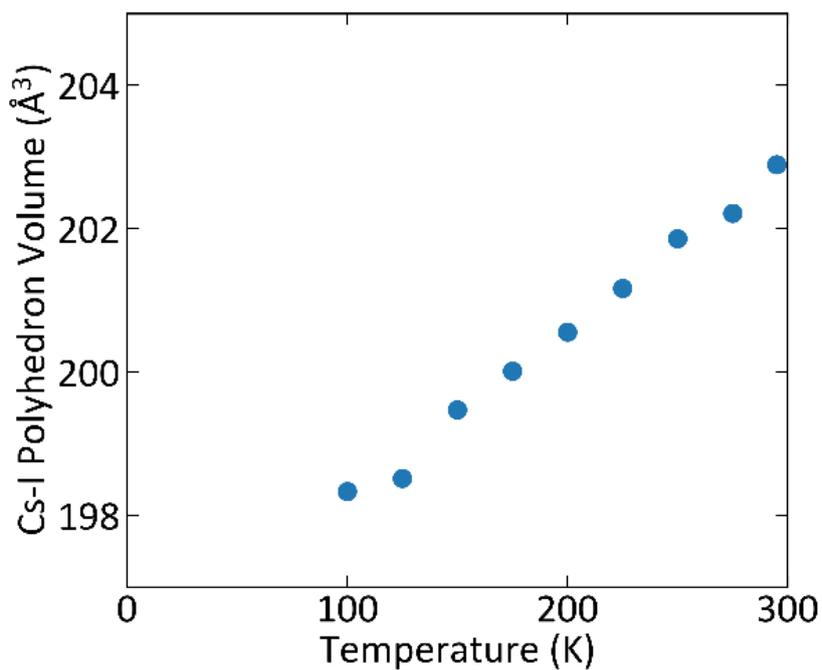

Figure S5: Cs-I polyhedron volume from SCXRD structures.

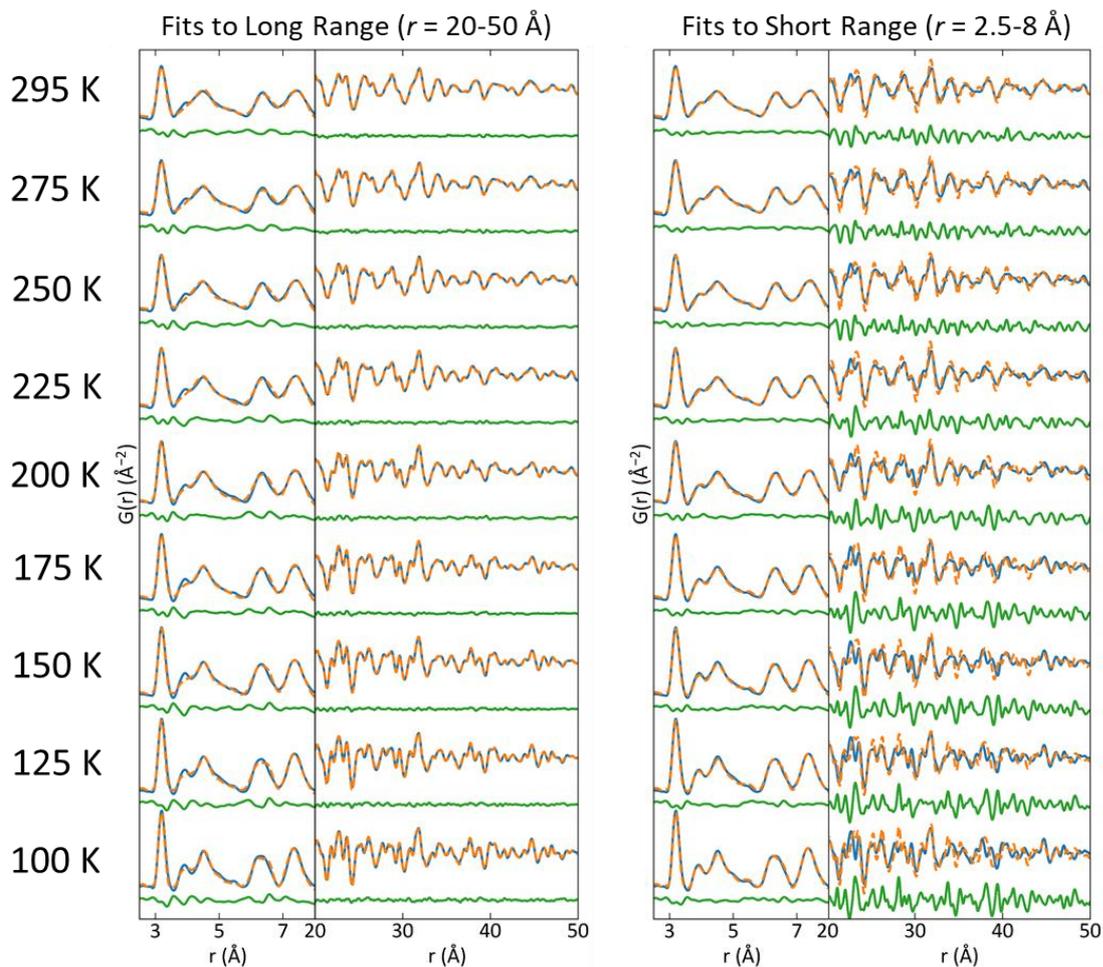

Figure S6: G(r) (blue) with fits (orange) and residual (green) to (left) long and (right) short range data.

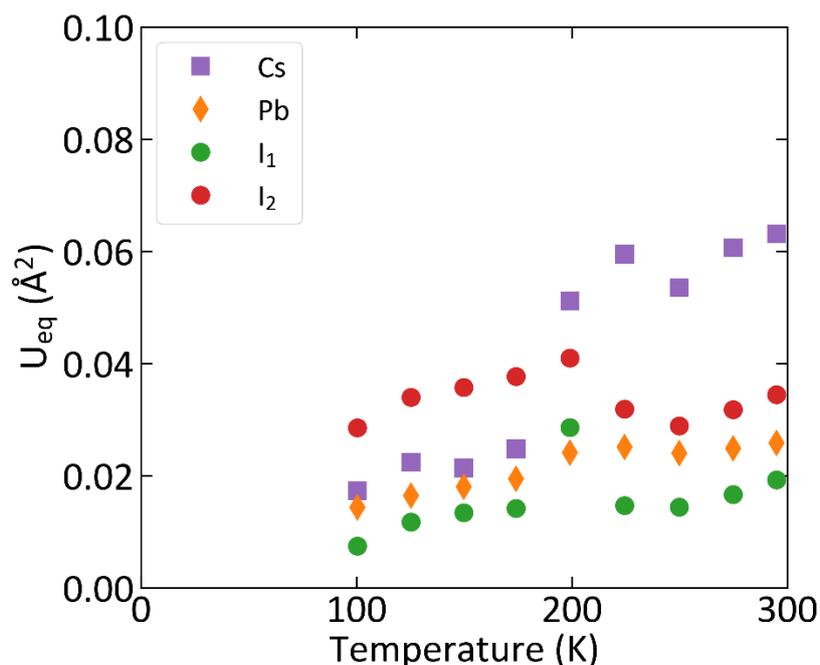

Figure S7: Equivalent isotropic thermal parameters ($U_{eq}$) from local PDF structures.

Table S1: Crystallographic parameters and structure refinement at 295 K

| | | |
|---|---|---|
| **Empirical formula** | $CsPbI_3$ | |
| **Formula weight** | 720.8 | |
| **Temperature/K** | 295.03 | |
| **Crystal system** | orthorhombic | |
| **Space group** | Pnma | |
| **a/Å** | 8.8637(8) | |
| **b/Å** | 12.4838(12) | |
| **c/Å** | 8.5783(8) | |
| **s, ß, ? /°** | 90 | |
| **Volume/Å3** | 949.21(15) | |
| **Z** | 4 | |
| **?calc g/cm3** | 5.044 | |
| **μ /mm-1** | 31.213 | |
| **F(000)** | 1184 | |
| **Crystal size/mm3** | 0.053 × 0.047 × 0.029 | |
| **Radiation** | MoKα (λ = 0.71073) | |
| **2θ range for data collection/°** | 5.762 to 56.564 | |
| **Index ranges** | -11 ≤ h ≤ 11, -16 ≤ k ≤ 16, -9 ≤ l ≤ 11 | |
| **Reflections collected** | 7193 | |
| **Independent reflections** | 1231 [Rint = 0.0518, Rsigma = 0.0458] | |
| | **Disordered Cs** | **One Cs** |
| **Data/restraints/parameters** | 1231/0/31 | 1231/0/28 |
| **Goodness-of-fit on F²** | 1.035 | 1.032 |
| **Final R indexes [I>=2σ (I)]** | R1 = 0.0318, wR2 = 0.0534 | R1 = 0.0344, wR2 = 0.0586 |
| **Final R indexes [all data]** | R1 = 0.0607, wR2 = 0.0605 | R1 = 0.0639, wR2 = 0.0665 |
| **Largest diff. peak/hole / e Å-3** | 1.24/-1.04 | 1.67/-1.76 |

Table S2: Crystallographic parameters and structure refinement at 100 K

| Empirical formula | $CsPbI_3$ | |
|---|---|---|
| Formula weight | 720.8 | |
| Temperature/K | 100.01 | |
| Crystal system | orthorhombic | |
| Space group | Pnma | |
| a/Å | 8.9586(8) | |
| b/Å | 12.4124(12) | |
| c/Å | 8.3383(8) | |
| σ, ß, γ /° | 90 | |
| Volume/Å$^3$ | 927.20(15) | |
| Z | 4 | |
| ρ$_{calc}$ g/cm3 | 5.164 | |
| μ /mm$^{-1}$ | 31.954 | |
| F(000) | 1184 | |
| Crystal size/mm$^3$ | 0.053 × 0.047 × 0.029 | |
| Radiation | MoKa (λ = 0.71073) | |
| 2θ range for data collection/° | 5.886 to 56.64 | |
| Index ranges | -11 = h = 11, -16 = k = 16, -9 = l = 11 | |
| Reflections collected | 7012 | |
| Independent reflections | 1203 [Rint = 0.0370, Rsigma = 0.0279] | |
| | **Disordered Cs** | **One Cs** |
| Data/restraints/parameters | 1203/0/31 | 1203/0/28 |
| Goodness-of-fit on F$^2$ | 0.989 | 0.99 |
| Final R indexes [I>=2σ (I)] | R1 = 0.0196, wR2 = 0.0353 | R1 = 0.0197, wR2 = 0.0355 |
| Final R indexes [all data] | R1 = 0.0303, wR2 = 0.0374 | R1 = 0.0304, wR2 = 0.0377 |
| Largest diff. peak/hole / e Å$^{-3}$ | 1.15/-1.07 | 1.16/-1.04 |